# Polar magnetism and chemical bond in α-RuCl$_3$

S. W. Lovesey[1, 2]

[1]ISIS Facility, STFC, Didcot, Oxfordshire OX11 0QX, UK

[2]Diamond Light Source Ltd, Didcot, Oxfordshire OX11 0DE, UK

**Abstract**

The micaceous black allotrope of ruthenium trichloride is the subject of many recent experimental and theoretical studies. Even so, its structural and magnetic properties remain undecided; monoclinic, trigonal and rhombohedral space groups for the crystal structure have been proposed on the basis of various types of experiments. The magnetic structure is often discussed in the context of the Kitaev state, but inevitably they are inconclusive discussions in the absence of structural and magnetic space groups. Johnson *et al*. infer a candidate for the magnetic structure (C$_c$2/m) from results gathered in an extensive set of experiments on an untwined sample of α-RuCl$_3$ [R. D. Johnson *et al*., Phys. Rev. B **92**, 235119 (2015)]. The proposed zigzag antiferromagnetic ground state of Ru ions does not respond to bulk magnetic probes, with optical rotation and all forms of dichroism prohibited by symmetry. Experimental techniques exploited by Johnson *et al*. included x-ray and magnetic neutron diffraction. Properties of the candidate magnetic structure not previously explored include polar magnetism that supports Ru Dirac multipoles, e.g., a ruthenium anapole that is also known as a toroidal dipole. In a general case, Dirac dipoles are capable of generating interactions between magnetic ions, as in an electrical Dzyaloshinskii-Moryia interaction [T. A. Kaplan and S. D. Mahanti, Phys. Rev. B **83**, 174432 (2011) and H. J. Zhao *et al*., Nat. Mat. **20**, 341 (2021)]. Notably, the existence of Dirac quadrupoles in the pseudo-gap phases of cuprate superconductors YBCO and Hg1201 account for observed magnetic Bragg diffraction patterns. Dirac multipoles contribute to the diffraction of both x-rays and neutrons, and a stringent test of the magnetic structure C$_c$2/m awaits future experiments. From symmetry-informed calculations we show that, the magnetic candidate permits Bragg spots that arise solely from Dirac multipoles. Stringent tests of C$_c$2/m can also be accomplished by performing resonant x-ray diffraction with signal enhancement from the chlorine K-edge. X-ray absorption spectra published for α-RuCl$_3$ possess a significant low-energy feature [K. W. Plumb *et al*., Phys. Rev. B **90**, 041112(R) (2014)]. Many experimental studies of other Cl-metal compounds concluded that identical features hallmark the chemical bond. Using a monoclinic C$_c$2/m structure, we predict the contribution to Bragg diffraction at the Cl K-edge absorption. Specifically, the variation of intensity of Bragg spots with rotation of the sample about the reflection vector. The two principal topics of our studies, polar magnetism and the chemical bond in the black allotrope of ruthenium trichloride, are brought together in a minimal model of magnetic Ru ions in C$_c$2/m.

## 1. Introduction

Ruthenium trichloride is one of the most widely used reagents [1], while the micaceous black allotrope α-RuCl$_3$ is much discussed in the context of a Kitaev state [2-6]. Signatures of the state have been reported on the basis of neutron scattering [7-9], nuclear magnetic resonance (NMR) [10], terahertz spectroscopy [11], microwave absorption [12], specific heat measurements [13], NMR spectroscopy [14], and Raman spectroscopy [15]. Discussions of experimental results are inconclusive in the absence of established space groups for both the chemical and magnetic structures of α-RuCl$_3$.

Johnson *et al.* infer a chemical structure C2/m and a magnetic structure C$_c$2/m from results gathered in an extensive set of experiments on an untwined sample of α-RuCl$_3$ [16]. Experimental techniques exploited by the authors included x-ray and magnetic neutron diffraction. We have studied the proposed magnetic structure with a view to identifying future experiments that can establish its validity beyond reasonable doubt. Stacking faults, magnetic exchange interactions, Ru-Ru bonds, band structure calculations etc. are discussed at length by Johnson *et al.*, and any further discussion of the topics is outside the scope of the present work. Properties of the candidate magnetic structure not previously explored include polar magnetism that supports Ru Dirac multipoles, which are both time-odd (magnetic) and parity-odd (polar). Best known, perhaps, is the Dirac dipole that is also called an anapole or toroidal dipole. Some contributions to magnetic scattering amplitudes contain no conventional (axial) magnetism. Evidently, such uniquely polar contributions to diffraction afford a stringent test of the C$_c$2/m magnetic structure. Enabling conditions in the halide mimic the first direct observation of Dirac multipoles by neutron diffraction [17]. The samarium compound in question adopts a diamond-like structure for which extensively studied (2, 2, 2) type Bragg reflections are basis-forbidden, i.e., a manifestation of angular anisotropy in the electron density. Strong intensities of the named Bragg spots from magnetic SmAl$_2$ were reliably attributed to Dirac multipoles in a symmetry-inspired calculation of the diffraction pattern. A parallel calculation of diffraction by magnetic α-RuCl$_3$ given here likewise appeals to basis-forbidden reflections, only this time from a magnetic motif on a monoclinic crystal structure for which reflections indexed ($h$, 0, $l$) with Miller indices $h$ even and $l$ odd are basis-forbidden [16].

Returning to α-RuCl$_3$ and diffraction patterns reported by Johnson *et al.*, two magnetic Bragg spots observed in difference data collected at temperatures of 6 K (magnetically ordered) and 20 K (paramagnetic), are indexed as (0, −2, −1) and (1, −1, −1) [16]. Relative intensity of the two spots could only be reproduced with conventional dipole moments confined to the plane normal to the unique b axis of the inferred monoclinic motif. As already mentioned, there is not a consensus on the appropriate space group for the chemical structure [7, 16, 18]. Lattice vibrations deduced from nuclear inelastic neutron scattering are essential for an interpretation of the half-integer thermal quantum Hall effect in α-RuCl$_3$ [7, 19].

Polar magnetism is not mentioned in previous studies of α-RuCl$_3$, to the best of our knowledge, and specifically not in Refs. [8, 16, 18]. Use of an untwined crystal and data refinement on a magnetic space group sets work by Johnson *et al.* apart [16]. Reported x-ray diffraction measurements display a single magnetic phase transition on cooling to low temperatures, in agreement with powder samples. A trigonal P3$_1$12 model for the crystal structure does not give a good fit to diffraction patterns. Later work by Cao *et al.* [18] agreed on the chemical structure reported in Ref. [16]. Two magnetic symmetries were found possible, Pm (zigzag) and Pm' (stripy), whereas Johnson *et al.* refine to C$_c$2/m. Ref. [18] does not contain relations between the mentioned space groups and measured magnetic intensities seem unrelated to reflections in C$_c$2/m that are created by polar magnetism. Cao *et al.* [18] mention the value of neutron polarization analysis in future studies of magnetic motifs, a technique epitomized in studies reported in Refs. [7, 20, 21]. Banerjee *et al*. [8] index magnetic reflections on the trigonal model without mention of conflicting results published in Ref. [16]. Notably, magnetic intensities they collected show two magnetic transitions.

The zigzag antiferromagnetic ground state of Ru ions in $C_c2/m$ does not present bulk magnetic properties. Bulk properties are controlled by the scattering amplitude, or structure factor, evaluated for the forward direction of scattering obtained with all Miller indices set to zero. In the case of $C_c2/m$, however, the Ru magnetic amplitude vanishes for $l$ even. The condition on the Miller index $l$ is a product of anti-translation in the magnetic unit cell, and the condition is made explicit in Eq. (1) for the Ru electronic structure factor. As for optical responses, polarization rotation and dichroic signals are prohibited in $C_c2/m$. These restrictions are violated in optical measurements on a shiny as-grown α-RuCl$_3$ sample [22]. Data for optical polarization rotation, also referred to by the authors as an equilibrium magneto-optical response or magnetic linear dichroism, are displayed in Fig. 2(b) of Ref. [22]. They are used to imply a phase transition in the sample at a temperature around 7 K. Linear dichroism is generated by electronic quadrupoles, be it axial or polar (non-reciprocal dichroism), e.g., electric dipole - electric dipole (E1-E1) or electric dipole - magnetic dipole (E1-M1) absorption events [23-26], and not the primary magnetic order-parameter utilized by the authors to interpret their data. Specifically, the assertion in Ref. [22] that the data reported in Fig. 2(b) establishes polarization spectroscopy as an all-optical alternative to magnetic neutron diffraction for probing a zizgzag magnetic ground state is not justified.

The second topic in our study is the chemical bond and what information about it is revealed by resonance enhanced x-ray Bragg diffraction. By way of orientation, we mention that numerous studies of x-ray absorption spectra (XAS) conclude that ligand K-edge spectroscopy is a direct probe of the covalency of a metal-ligand bond [27-30]. An x-ray absorption profile for α-RuCl$_3$ displays a bold feature at the chlorine K-edge; see Fig. S2 in Ref. [4]. Very informative absorption spectra are those of isostructural structures hosting [ZnCl$_4$] and [CuCl$_4$] units that exhibit major activities at an energy ≈ 2827 eV [28]. Additionally, the copper but not the zinc compound shows an intense low energy peak at ≈ 2820 eV. As Zn$^{2+}$ has a filled d$^{10}$ shell, the lowest unoccupied energy levels are the 4s and 4p orbitals. Thus, its edge features derive primarily from the E1 1s → 4p transition. Whereas, Cu$^{2+}$ possesses a d$^9$ configuration with a hole in the d shell. The observed pre-edge transition was assigned by the authors as a Cl 1s → Cu 3d transition, not possible in the Zn compound because of the filled d subshell [28]. This assignment appeals to a covalent bond between Cu 3d and Cl 3p orbitals, giving rise to a half-occupied antibonding wave function. If ρ$^2$, say, represents the Cl 3p character in the molecular orbital the same quantity controls the intensity of a Cl-centred 1s → 3p E1 transition. Likewise, chlorine multipoles observed in resonant x-ray diffraction depend on the covalency parameter. A direct Cl 1s → Cu 3d transition is accomplished with an electric-quadrupole (E2) event that has a relative amplitude $(1 - ρ^2)^{1/2}$. The Cl 3p-metal d hybridization leads to holes in the Cl p-orbital and E1-E1 events in x-ray scattering. We report amplitudes for x-ray Bragg diffraction based on the monoclinic structure of α-RuCl$_3$ that include rotation of the crystal about the reflection vector (an azimuthal angle scan).

The two topics of study, polar magnetism and the chemical bond, are brought together in Section IV, where a minimal model for the Ru atomic state is introduced. The model includes two atomic manifolds belonging to total angular momenta j = 3/2 and j′ = 5/2. Thereby, we account for the rare observation of strong x-ray absorption at the $L_2$ edge [4]. Moreover, an axial quadrupole in magnetic neutron scattering is non-zero with two j-manifolds in the Ru atomic state, and we derive an explicit expression from the minimal model. Our result for the axial dipole contribution to the magnetic neutron scattering amplitude includes significant orbital angular momentum.

## 2. Magnetic structure

The parent crystal structure of untwined α-RuCl$_3$ C2/m (No.12) is monoclinic and non-centrosymmetric, and depicted in Fig. 1. The unique axis b is parallel to a dyad axis of rotation symmetry (2ζ) in sites 4g used by ruthenium ions. Chlorine ions are in 4i and 8j, and the latter have no symmetry whereas sites 4i possess mirror symmetry mζ = Ī2ζ. Obtuse cell vectors a(sin(β), 0, cos(β)), (0, 0, c) and unique (0, b, 0), with cell lengths a ≈ 5.9762 Å, b ≈ 10.3420 Å, c ≈ 6.0130 Å, and β ≈ 108.87° [16].

Ruthenium ions (Ru$^{3+}$, 4d$^5$), undergo long-range antiferromagnetic order at a low temperature ≈ 13 K. A magnetic space group C$_c$2/m (BNS setting No. 12.63) describes the order inferred from neutron powder diffraction data gathered deep in the ordered phase (6 K) [16], with Ru ions in sites 8h and constrained by symmetry 2′ζ. Antiferromagnets with a nontrivial magnetic Bravais lattice have a crystal class containing 1′, and the magnetic crystal class 2/m1′ of C$_c$2/m is one of 11 grey classes that contain all three inversions $\bar{1}$, 1′ and $\bar{1}$′. Any type of magnetoelectric effect is prohibited, likewise optical rotation. Basis vectors relative to the parent C2/m structure are {(1, 0, 2), (0, −1, 0), (0, 0, −2)}, and they are denoted (**a**$_m$, **b**$_m$, **c**$_m$) with **b**$_m$ labelled ζ. The obtuse angle = cos$^{-1}${− (cos(β) + 2r)/√[4r (cos(β) + r) +1]} ≈ 150.74° with r = c/a. The reciprocal lattice has volume v$_o$ = (2 abc) sin(β), and vectors **a**$_m$* = (2π/v$_o$) (2 bc, 0, 0), **b**$_m$* =(2π/v$_o$) (0, −2 ac sin(β), 0) **c**$_m$* = (2π/v$_o$) (b (a cos(β) + 2 c), 0, − ab sin(β)).

Electronic properties of Ru and Cl ions are encapsulated in spherical multipoles ⟨O$^K_Q$⟩ that have integer rank K and projections Q that obey − K ≤ Q ≤ K [23, 31] (Cartesian and spherical components of a dipole **R** = (x, y, z) are related by x = (R$_{-1}$ − R$_{+1}$)/√2, y = i(R$_{-1}$ + R$_{+1}$)/√2, z = R$_0$). A complex conjugate is defined as ⟨O$^K_Q$⟩* = (−1)$^Q$ ⟨O$^K_{-Q}$⟩, and a phase convention ⟨O$^K_Q$⟩ = [⟨O$^K_Q$⟩′ + i⟨O$^K_Q$⟩″] for real and imaginary parts labelled by single and double primes, respectively. Angular brackets ⟨ ... ⟩ denote the time-average, or expectation value, of the enclosed tensor operator.

## 3. Neutron diffraction

An electronic structure factor Ψ$^K_Q$ = [exp(i**κ** • **d**) ⟨O$^K_Q$⟩**d**], where the reflection vector **κ** is defined by integer Miller indices (h, k, l), and the implied sum is over all ions in a unit cell. Relations between Miller indices for the parent structure (H$_o$, K$_o$, L$_o$) and (h, k, l) for the magnetic motif of Ru ions in C$_c$2/m are h = H$_o$ + 2L$_o$, k = −K$_o$, l = − 2L$_o$. Orthonormal Cartesian coordinates ($\xi$, $\eta$, $\zeta$) are derived from ($c_m$, $a_m^*$, $b_m$). Components of a unit reflection vector $\kappa_\xi = -(l/2\kappa c)$, $\kappa_\eta = [(2/a)(h+l) + (l/c)\cos(\beta)]/[2\kappa\sin(\beta)]$, and $\kappa_\zeta = -(k/\kappa b)$.

An anti-dyad $2'\zeta$ imposes the constraint Q odd on projections of Ru magnetic multipoles. Dipoles (K = 1) are parallel to $\xi$- and $\eta$-axes. The Ru axial quadrupole is different from zero when the atomic wave function is drawn from two, or more, j-manifolds, cf. Eq. (11). Such is the case in the $j_{eff} = 1/2$ state of Ir ions ($Ir^{4+}$, $5d^5$), often exploited in studies of $Sr_2IrO_4$, which is an admixture of manifolds $j = 3/2$ and $j = 5/2$ [32, 33].

Returning to Ru ions in $C_c2/m$,

$$\Psi^K{}_Q(8h) = [1 + \sigma_\theta (-1)^l] [1 + (-1)^{h+k}]$$
$$\times \langle O^K{}_Q \rangle \exp(i\pi\, l/2) [\exp(i\varphi) + \sigma_\pi \sigma_\theta \exp(-i\varphi)], \tag{1}$$

where the spatial phase factor $\varphi = (2\pi y k)$ and the fractional coordinate $y \approx 0.166$ [16]. A parity $\sigma_\pi$ (time $\sigma_\theta$) signature $\sigma_\pi = +1$ (−1) for axial (polar) multipoles while $\sigma_\theta = -1$ for a magnetic ion. Conventional magnetism is axial, meaning $\sigma_\pi \sigma_\theta = -1$, while Dirac multipoles are polar and $\sigma_\pi \sigma_\theta = +1$.

Bulk magnetic signals are not permitted, namely, $\Psi^K{}_Q(8h) = 0$ for $l = 0$ and $\sigma_\theta = -1$. The mentioned condition is provided by the first factor in Eq. (1) that arises from anti-translation in the unit cell. Since the condition is independent of the tensor rank K and parity $\sigma_\pi$ it rules against all types optical and x-ray dichroic signals, for example, from the zigzag antiferromagnetic state of Ru ions.

In Eq. (1) we observe reflection conditions $h + k$ even (C-centring) and $l$ odd (anti-translation in $C_c2/m$). Setting $\sigma_\theta = -1$,

$$\Psi^K{}_Q(8h) = 4 \langle O^K{}_Q \rangle \exp(i\pi\, l/2) [\exp(i\varphi) - \sigma_\pi \exp(-i\varphi)]. \tag{2}$$

Evidently, Bragg spots indexed by $k = \varphi = 0$ are due to polar magnetism alone.

First, though, more familiar axial contributions to a neutron magnetic scattering amplitude,

$$\langle Q_\xi \rangle^{(+)} \approx (3/2) \langle T^1{}_\xi \rangle - \sqrt{3}[\kappa_\xi \kappa_\eta \langle T^2{}_{+1} \rangle' + (\kappa_\eta{}^2 - \kappa_\zeta{}^2) \langle T^2{}_{+1} \rangle''], \tag{3}$$

$$\langle Q_\eta \rangle^{(+)} \approx (3/2) \langle T^1{}_\eta \rangle + \sqrt{3}[\kappa_\xi \kappa_\eta \langle T^2{}_{+1} \rangle'' + (\kappa_\xi{}^2 - \kappa_\zeta{}^2) \langle T^2{}_{+1} \rangle'],$$

$$\langle Q_\zeta \rangle^{(+)} \approx \sqrt{3}\, \kappa_\zeta (\kappa_\eta \langle T^2{}_{+1} \rangle' - \kappa_\xi \langle T^2{}_{+1} \rangle''). \qquad k \text{ non-zero}, l = (2n + 1)$$

For the sake of clarity, we omit a common factor $\{-8(-1)^n \sin(\varphi)\}$ in Eq. (3) that vanishes for $k = 0$. The dipole,

$$\langle T^1 \rangle \approx (1/3) \{2\langle S \rangle \langle j_0(\kappa) \rangle + \langle L \rangle [\langle j_0(\kappa) \rangle + \langle j_2(\kappa) \rangle]\}, \tag{4}$$

to a good approximation [31]. Standard radial integrals in Eq. (4) satisfy $\langle j_0(0) \rangle = 1$ and $\langle j_2(0) \rangle = 0$ [34]. In consequence, $\langle T^1 \rangle$ approaches $\mu_o/3$ in the forward direction $\kappa = 0$, where $\mu_o = [2\langle S \rangle$

+ ⟨**L**⟩] is the Ru magnetic moment. The axial quadrupole in Eq. (3) is proportional to ⟨$j_2(\kappa)$⟩ and more is said about it in Section 4 [35, 36]. Intensity of a Bragg spot = |⟨**Q**⊥⟩|$^2$ with ⟨**Q**⊥⟩$^{(+)}$ = [**κ** × (⟨**Q**⟩$^{(+)}$ × **κ**)]/$\kappa^2$.

A Dirac dipole ⟨**D**⟩ is the sum of three contributions that include spin and orbital Ru anapoles, also known as toroidal dipoles Fig. 2, and each contribution accompanied by a unique dependence on the magnitude of the reflection vector, i.e., there are three atomic form factors in ⟨**D**⟩, which have been calculated for several atomic configurations [37, 38]. Operators for the three contributions are a spin anapole (**S**×**R**), orbital anapole **Ω** = [**L**×**R** − **R**×**L**], and (i**R**). The foregoing dipole ⟨**T**$^1$⟩ shows that intensities of axial Bragg spots depend on spin and orbital angular magnetizations, and the latter is bound to be significant, as illustrated in Section IV. The total dipole scattering amplitude is the sum of ⟨**Q**⊥⟩$^{(+)}$ and Dirac dipole contributions,

$$⟨Q_{⊥ξ}⟩^{(-)} ≈ - iκ_ζ ⟨D_η⟩/κ, ⟨Q_{⊥η}⟩^{(-)} ≈ iκ_ζ ⟨D_ξ⟩/κ, ⟨Q_{⊥ζ}⟩^{(-)} ≈ i[κ_ξ ⟨D_η⟩ - κ_η ⟨D_ξ⟩]/κ. \quad (5)$$

The common factor in ⟨**Q**⊥⟩$^{(-)}$ is {8i (−1)$^n$ cos(φ)}. Intensity of Bragg spots indexed (*h*, 0, *l*) with even *h* and odd *l* is due solely to polar magnetism, expressed by anapoles in Eq. (5) at the first level of completeness. Dipole contributions ⟨$Q_{⊥ξ}$⟩$^{(-)}$ = ⟨$Q_{⊥ζ}$⟩$^{(-)}$ = 0 and,

$$|⟨\mathbf{Q}_⊥⟩^{(-)}|^2 = (8/κ)^2 |(κ_ξ ⟨D_η⟩ - κ_η ⟨D_ξ⟩)|^2 \text{ for } (h, 0, l), \quad (6)$$

follows from Eq. (5). Recall that $κ_η ∝ [(2/a) (h + l) + (l/c) \cos(β)]$ and $κ_ξ ∝ (l/c)$. Published diffraction data is for Bragg spots (0, −2, −1) and (1, −1, −1) should properly be interpreted with an intensity |⟨**Q**⊥⟩$^{(+)}$ + ⟨**Q**⊥⟩$^{(-)}$|$^2$.

## 4. Minimal model

The crystal field experienced by a Ru ion is expected to dominate the spin coupling in α-RuCl$_3$ resulting in a low-spin d$^5$ configuration [39]. All four $e_g$ states are occupied and one $t_{2g}$ state is unoccupied. The orbital triplet carries spin S = 1/2 and a magnetic moment √[S (S + 1)] = 1.73. A realistic Ru wave function must include the j = 3/2 manifold to ensure that x-ray absorption at the L$_2$ edge is allowed [4]. Not so for Sr$_2$IrO$_4$ with no intensity reported at the L$_2$ edge [32, 33]. A lower limit for the Ru saturation magnetic moment $μ_o$ ≈ 0.64(4) $μ_B$ has been reported [16].

With these facts in mind, we explore axial magnetic multipoles derived from a Kramers doublet composed of |u⟩ and its time-reversed partner |û⟩. A ground state |g⟩ = [|u⟩ + f |û⟩]/√[1 + |f|$^2$] complies with site symmetry 2′$_ζ$ for a mixing parameter |f|$^2$ = 1 [33]. We find a minimal state,

$$|u⟩ = N_o [α |j, m⟩ + β |j', m'⟩], \quad (7)$$

with real α, complex β and normalization $N_o^2 [α^2 + |β|^2] = 1$ yields plausible multipoles on adopting values j = 3/2, m = 1/2, j′ = 5/2, m′ = −3/2, i.e., Δm = ±2 compatible with a dyad. A saturation magnetic moment is derived from the result,

$$⟨μ_{+1}⟩ = ⟨g|μ_{+1}|g⟩ = √2 (1/5) N_o^2 α^2 [2f + (β/α) f^* √3], \quad (8)$$

using $\mu_o = \sqrt{[|\langle\mu_{+1}\rangle|^2 + |\langle\mu_{-1}\rangle|^2]}$. And $\langle\mu_\xi\rangle = -\sqrt{2}\,\langle\mu_{+1}\rangle'$, $\langle\mu_\eta\rangle = -\sqrt{2}\,\langle\mu_{+1}\rangle''$ for magnetic dipole moments in the plane normal to the unique axis b. For the extreme case $\langle\mu_\eta\rangle = 0$, achieved with real β and f = 1, the observed $\mu_o = 0.64$ occurs for (β/α) = − 0.20. A negative ratio of coefficients increases orbital angular momentum,

$$\langle L_{+1}\rangle = (1/5\sqrt{2})\,N_o^2\,\alpha^2\,[8f - (\beta/\alpha)\,f^*\,\sqrt{3}]. \tag{9}$$

For reference, $\langle L_{+1}\rangle/\langle\mu_{+1}\rangle = [(2 - g_o)/g_o]$ and $\langle S_{+1}\rangle/\langle\mu_{+1}\rangle = [(g_o - 1)/g_o]$ in the absence of |j', m'⟩, where the Landé factor $g_o = 4/5$ for j = 3/2.

The Ru axial dipole in the magnetic neutron scattering amplitude Eq. (3) assumes the value,

$$\langle T^1_{+1}\rangle = \sqrt{2}\,(1/15)\,N_o^2\,\alpha^2\,\{2f\,[\langle j_0(\kappa)\rangle + 2\,\langle j_2(\kappa)\rangle] + (\beta/\alpha)\,f^*\,\sqrt{3}[\langle j_0(\kappa)\rangle - (1/2)\,\langle j_2(\kappa)\rangle]\}. \tag{10}$$

Here, we use exact results for reduced matrix elements for states j = 3/2, j′ = 5/2 [31]. A negative (β/α) enhances the contribution to $\langle T^1\rangle$ of the radial integral $\langle j_2(\kappa)\rangle$ which must be included in reliable data analysis. Adopting real β, f = 1 and (β/α) = − 0.20, a conventional atomic form factor for Ru ions becomes $[\langle j_0(\kappa)\rangle + 2.523\,\langle j_2(\kappa)\rangle]$. Values for the Bragg reflection (0, −2, −1) at κ ≈ 1.61 Å$^{-1}$ yield $(2.523\,\langle j_2(\kappa)\rangle/\langle j_0(\kappa)\rangle) \approx 0.30$ (d-spacing = (2π)/κ ≈ 3.90 Å). The first maximum $\langle j_2(\kappa)\rangle \approx 0.23$ is at κ ≈ 4.06 Å$^{-1}$, with $\langle j_0(\kappa)\rangle \approx 0.17$.

Next in line to the dipole is a quadrupole for which we find,

$$\langle T^2_{+1}\rangle = -i\,(5/42)\,N_o^2\,\alpha\,\beta\,f^*\,\langle j_2(\kappa)\rangle. \tag{11}$$

This result demonstrates the general condition that quadrupoles in magnetic neutron scattering arise from an admixture of manifolds with different total angular momenta. The equivalent operator [(**S**×**R**) **R**] for **T**$^2$ shows that actually the quadrupole measures the correlation between the spin anapole and orbital degrees of freedom [31, 36]. In the extreme case where parameters are purely real, and the axial dipole is aligned with the ξ axis, the quadrupole is purely imaginary.

The low-energy electronic degrees of freedom are primarily pdσ antibonding Cl p-type and metal d-type orbitals. The same anion and cation states determine the hybridization matrix that is proportional to the wavefunction overlap of metal and Cl holes. Chlorine 3p electrons contribute to the ruthenium ground state and polar multipoles. Indeed, in the vicinity of a Ru ion chlorine 3p wave functions can be expanded in the basis of the Ru orbitals [40]. Specifically, Cl 3p penetrate Ru 5p orbitals and give a contribution to anapoles in the Dirac dipole ⟨**D**⟩. Thereby, a Kramers state [|u⟩ + f |û⟩] receives an addition λ|5p⟩ and $\langle \mathbf{G}^K\rangle \propto \lambda$ [⟨u|**G**$^K$|5p⟩ + ⟨5p|**G**$^K$|u⟩ + f ⟨5p|**G**$^K$|û⟩ + f* ⟨û|**G**$^K$|5p⟩], where ⟨**G**$^K$⟩ is a Dirac multipole of rank K and **G**$^1$ = **D**. Quadrupoles ⟨**G**$^2$⟩ describe magnetization in pseudo-gap phases of cuprate superconductors YBCO and Hg1201 uncovered with neutron polarization analysis [41, 42].

Specifically, ⟨**D**⟩ = (1/2) [3 (h$_1$) ⟨(**S**×**R**)⟩ − (j$_0$) ⟨**Ω**⟩ + (g$_1$) ⟨(i**R**)⟩]. Values of the atomic form factors for the Bragg reflection (0, −2, −1) are (h$_1$) ≈ 0.13, (j$_0$) ≈ − 0.09 and (g$_1$) ≈ 1.00 indicate a requirement to include anapoles in the analysis of a diffraction pattern. For reference, form factors for the axial dipole have values $\langle j_0(\kappa)\rangle \approx 0.76$ and $\langle j_2(\kappa)\rangle \approx 0.09$ at $(0, -2, -1)$. Notably, polar magnetism alone provides magnetic intensity at reflections $(h, 0, l)$.

**5. Resonant x-ray Bragg diffraction**

Valence states that accept the photo-ejected electron interact with neighbouring ions. In consequence, any corresponding electronic multipole is rotationally anisotropic with a symmetry corresponding to the site symmetry of the resonant ion [23, 43 – 45]. Tuning the energy of x-rays to an atomic resonance has two obvious benefits in diffraction experiments of interest; first, enhancement of Bragg spot intensities and, secondly, spots are element specific. In addition, there are four scattering amplitudes labelled by photon polarization, two with unrotated and two with rotated states of polarization [43, 44]. Amplitudes for rotated states of polarization exclude isotropic charge that presents strong Thomson scattering that overwhelms weak signals caused by charge anisotropy. The presence of strong scattering in unrotated channels of polarization is correct for scattering enhanced by a parity-even absorption event, e.g., electric dipole - electric dipole (E1-E1), but not so for a parity-odd absorption. We consider the electric dipole - electric quadrupole (E1-E2) event. All amplitudes calculated for E1-E1 and E1-E2 events include rotation of the crystal about the reflection vector (azimuthal-angle scan) and they are specific to position multiplicity, Wyckoff letter and symmetry. Axial and polar multipoles are denoted $\langle \mathbf{t}^K\rangle$, and $\langle \mathbf{u}^K\rangle$ and $\langle \mathbf{g}^K\rangle$, respectively. The range of values of the rank K is fixed by the triangle rule, and K = 0 – 2 and K = 1 – 3 for E1-E1 and E1-E2 events, respectively. The scalar $\langle \mathbf{t}^0\rangle$ is responsible for strong Thomson scattering in conventional, non-resonant x-ray Bragg diffraction. Angular rotation symmetry in a crystal can be mirrored in periodicity of an azimuthal-angle scan, and the valuable property of the diffraction technique depends on the direction of the chosen reflection vector.

A conventional labelling of photon polarization states is $\sigma$ and $\pi$ for primary polarization perpendicular and parallel to the plane of scattering, and primed labels for states of secondary polarization [23, 43, 46]. All required diffraction amplitudes using E1-E1 and E1-E2 absorption events are listed by Scagnoli and Lovesey [46]. They are functions of the Bragg angle $\theta$, azimuthal angle $\psi$, and polarization states.

Chlorine ions use sites 8i and 16j in $C_c2/m$ [16]. The sites are not equivalent and Bragg intensities add. The energy of the chlorine K-edge equates to a photon wavelength $\approx 4.40$ Å and the Laue condition for Bragg diffraction can be satisfied for a small range of Miller indices. X-ray amplitudes for a reflection $(0, k, 0)$ are proportional to spatial factors of the type $\cos(2\pi y k)$ or $\sin(2\pi y k)$ for axial ($\sigma_\pi = +1$) and polar ($\sigma_\pi = -1$) chlorine multipoles, respectively, as in foregoing amplitudes for neutron diffraction. The result $\sin(\theta) \approx (0.213\, k)$ shows that Bragg spots $k = 2$ and 4 are available. The general coordinate y = 0 for chlorine sites 8i, and the E1-E1 amplitude in the rotated channel of polarization is,

$$F_{\pi'\sigma}(8i) = 4 \sin(\theta) \{\sin(2\psi) [-\sqrt{(3/2)} \langle t^2_0\rangle + \langle t^2_{+2}\rangle'] + \cos(2\psi) \langle t^2_{+1}\rangle' \}. \quad (0, k, 0) \quad (12)$$

The two-fold rotation symmetry in the azimuthal angle stems from the dyad on unique axis b, and $F_{\pi'\sigma}(16j) = 2 \cos(2\pi y k)\, F_{\pi'\sigma}(8i)$ with $y \approx 0.327$ [16]. The crystal axis $\mathbf{c}_m$ is normal to the plane of scattering at the origin $\psi = 0$. There is no intensity at $(0, k, 0)$ from polar multipoles in 8i, while the amplitude for Cl ions in 16j sites is proportional to $\sin(2\pi y k) = -0.820$ and 0.939 for $k = 2$ and 4, respectively. On these grounds alone, polar and axial Bragg spot intensities for sites 16j are in a ratio $\approx 7.4$ for $k = 4$. Dipoles do not contribute to the E1-E2 amplitude in the rotated channel of polarization,

$$F_{\pi'\sigma}(16j) \approx -2\sqrt{(2/15)} \sin(2\pi y k) \{\sqrt{(3/2)} [\Phi \cos(2\psi) - 2\cos^2(\theta)] \langle u^2_0 \rangle \quad (0, k, 0)$$

$$- [\Phi \cos(2\psi) + 6\cos^2(\theta)] \langle u^2_{+2} \rangle' + 2\Phi \sin(2\psi) \langle u^2_{+1} \rangle' \}, \quad (13)$$

where $\Phi = [5\cos(2\theta) + 1]$, and chlorine polar octupoles $\langle \mathbf{u}^3 \rangle$ are omitted for simplicity. However, the allowed chlorine dipole $\langle u^1_{+1} \rangle''$ contributes to both unrotated amplitudes where the contributions do not depend on $\psi$.

Examination of the experimentally determined general coordinates for chlorine ions reveals that E1-E1 enhanced diffraction indexed $(h, 0, 0)$ with $h = 2$ is strong, since $\cos(2\pi x h) \approx 1.0$ for 8i ($x \approx 0.230$) and 16j ($x \approx 0.751$); conversely, E1-E2 enhanced diffraction is negligible [8]. The amplitude for E1-E1 diffraction in the rotated channel of polarization is,

$$F_{\pi'\sigma}(8i) = -4\cos(2\pi x h) \{\sin(\theta) \sin(2\psi) [\sqrt{(3/2)} \langle t^2_0 \rangle + \langle t^2_{+2} \rangle'] \quad (h, 0, 0)$$

$$+ \cos(\theta) \cos(\psi) \langle t^2_{+1} \rangle' \}, \quad (14)$$

with the same result for $F_{\pi'\sigma}(16j)$ apart from an overall factor 2 and the appropriate fractional coordinate.

Ruthenium anapoles contribute to the amplitude for resonant x-ray diffraction, together with Dirac multipoles $\langle \mathbf{g}^K \rangle$ of higher rank $K > 1$. By way of an example, consider a basis-forbidden Bragg spot $(0, 0, l)$. Exploiting the Ru $L_2$ or $L_3$ absorption edges permits $l = 1$, for $\sin(\theta) \approx (0.374\, l)$. Anapoles and quadrupoles satisfy sum rules $[\langle \mathbf{g}^1 \rangle_{L2} + \langle \mathbf{g}^1 \rangle_{L3}] \propto \langle \mathbf{\Omega} \rangle$ and $[\langle \mathbf{g}^2 \rangle_{L2} + \langle \mathbf{g}^2 \rangle_{L3}] \propto \langle \{\mathbf{L} \otimes \mathbf{R}\}^2 \rangle$, where in the latter we use a standard tensor product. The E1-E2 amplitude for rotated polarization that is correct for $K = 1, 2$ and $l = 1$ [46] is,

$$F_{\pi'\sigma}(8h) \approx (3/\sqrt{10}) \sin(2\theta) [-\sin(\chi) \cos(\psi) \langle g^1_\xi \rangle + \sin(\psi) \langle g^1_\eta \rangle]$$

$$+ \langle g^2_{+1} \rangle' [\sin(2\chi) (2\cos(2\theta) + \cos^2(\theta) \cos^2(\psi) - 1) + 2\cos(2\chi) \sin(2\theta) \sin(\psi)]$$

$$- 2 \langle g^2_{+1} \rangle'' \cos(\theta) \cos(\psi) [\cos(\chi) \cos(\theta) \sin(\psi) + 2\sin(\chi) \sin(\theta)], \quad (15)$$

omitting a common factor $4i \sqrt{(2/15)}$, and $\chi \approx -29.261°$. The unique axis b is in the plane of scattering for $\psi = 0$. Intensity $|F_{\pi'\sigma}(8h)|^2$ depends on the sign of $l$ when quadrupoles are present.

## 6. Discussion

In summary, we have revisited the magnetic structure and chemical bond in the allotrope $\alpha$-RuCl$_3$ taking on board experimental results obtained with neutrons and x-rays [4, 16]. An objective of our work is to stimulate further studies using these radiations. For one thing, the contribution of polar magnetism is an untouched question, although its existence is inferred by currently available experimental literature.

There is extensive literature on investigations of chlorine-metal covalent bonding using x-ray absorption, but not so for α-RuCl$_3$, to the best of our knowledge. To this end, we have calculated amplitudes suitable for diffraction enhance by the Cl K-edge, which include the dependence of Bragg spot intensities on rotation of the crystal about the reflection vector (azimuthal scan).

Compulsory polar magnetism in the zigzag antiferromagnetic ground state of α-RuCl$_3$ supports Dirac (magnetoelectric) multipoles [16]. Direct evidence of Dirac multipoles is available from magnetic neutron and resonant x-ray diffraction, and representative scattering amplitudes from symmetry-informed calculations are Eq. (5) and Eq. (15), respectively. At the first level, they contribute three types of anapoles each with an individual atomic form factor. The extreme sensitivity offered by neutron polarization analysis has revealed magnetization in the pseudo-gap phases of high T$_c$ materials, and anapoles in magnetic materials with a diamond-type structure [17, 20, 21]. Johnson *et al.* [16] relied on a difference of patterns taken in the paramagnetic (20 K) and ordered magnetic (6K) phases. In either method for isolating the magnetic content in the intensity of scattered neutrons, it is necessary to make adjustment for the atomic form factor. Especially for α-RuCl$_3$ with a large Ru orbital moment.

There are many reports of studies of Dirac multipoles using resonant x-ray Bragg diffraction, e.g., V$_2$O$_3$ [47, 48], GaFeO$_3$ [49] and α-Fe$_2$O$_3$ [50]. We calculated contributions made by Ru anapoles and quadrupoles to the x-ray diffraction amplitude as examples of the wealth of information that is potentially available in future experiments on α-RuCl$_3$. Values for the essential spin-orbit coupling can be derived from branching ratios [51, 52], with potentially useful data published by Plumb *et al.* [4], see Fig. 2(c) therein. Bulk magnetic signals are not permitted from the Ru zigzag antiferromagnetic ground state. This finding is discussed in light of experimental data for optical polarization rotation as a function of temperature reported for a shiny as-grown α-RuCl$_3$ sample [22].

**Acknowledgement** Dr R. D. Johnson specified the α-RuCl$_3$ magnetic motif inferred from neutron powder diffraction, provided Fig. 1, and debated findings reported in the paper. Dr D. D. Khalyavin provided valuable insight to the symmetry of the magnetic structure. Dr V. Scagnoli provided Fig. 2, and Professor G. van der Laan provided radial integrals and valuable insights on dichroic signals.

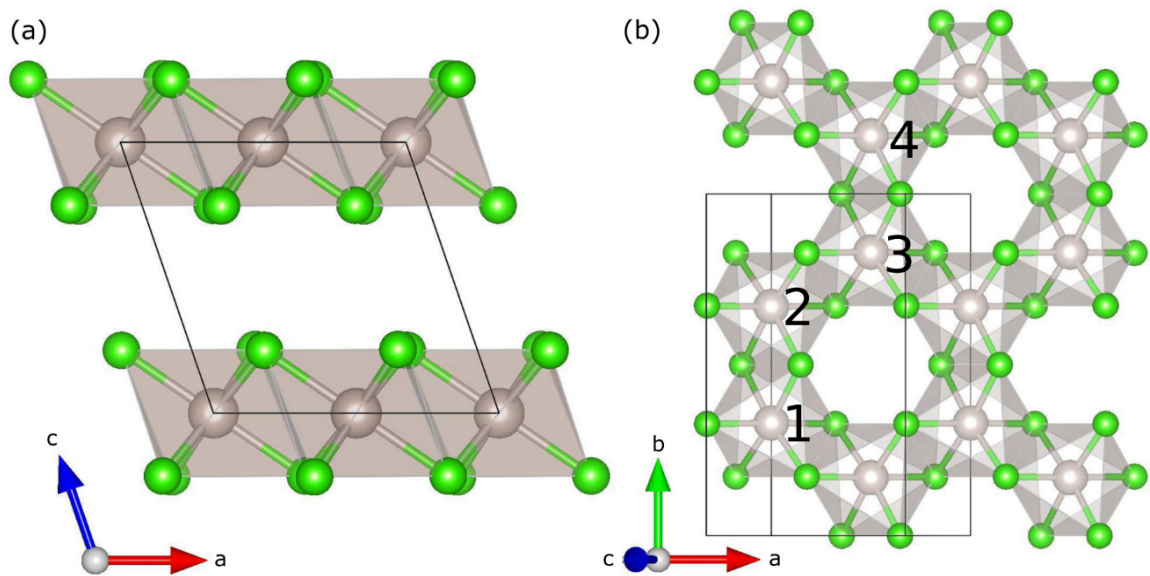

**Figure 1**. Honeycomb layers of edge-sharing RuCl$_6$ octahedra stacked vertically with an in-plane offset. Monoclinic α-RuCl$_3$ unit cell in black outline, and Ru and Cl ions grey and green spheres, respectively. (a) Projection onto the ac plane. (b) Basal layer projected onto the ab plane. Numbers in panel (b) label sites 4g in C2/m used by Ru ions. After Ref. [16].

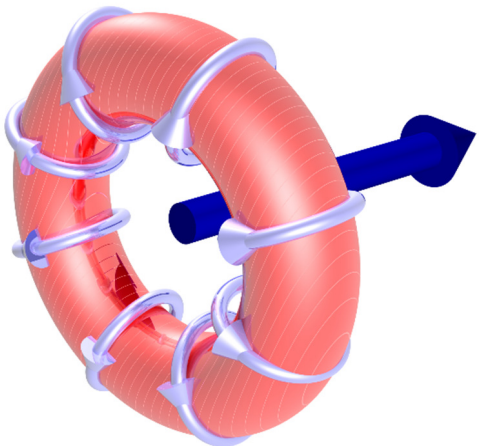

**Figure 2**. Depiction of a toroidal dipole, also known as an anapole.